\documentclass[a4paper,fleqn,usenatbib]{mnras}
\usepackage{newtxtext,newtxmath}
\usepackage[T1]{fontenc}
\usepackage{ae,aecompl}
\usepackage{graphicx}	
\usepackage{amsmath}	
\usepackage{amssymb}	
\usepackage{dblfloatfix}
\usepackage{color,soul}
\usepackage{indentfirst}
\usepackage{url}
\usepackage{scalerel}
\usepackage{placeins}
\usepackage{subfig}
\usepackage[textwidth=1.5cm,shadow]{todonotes}
\usepackage[normalem]{ulem} 
\usepackage{lineno}
\def\code#1{\texttt{#1}}

\title[Formation of MSPs in $\gamma$-ray bright globular clusters]{Milky Way globular clusters in $\gamma$-rays: analyzing the dynamical formation of millisecond pulsars}

\author[de Menezes et al.]{
Raniere de Menezes,$^{1}$\thanks{E-mail: raniere.m.menezes@gmail.com}
Fabio Cafardo,$^{1}$ 
Rodrigo Nemmen$^{1}$
\\
$^{1}$Universidade de S\~ao Paulo, Instituto de Astronomia, Geof\'{\i}sica e Ci\^encias Atmosf\'ericas, Departamento de Astronomia,\\ S\~ao Paulo, SP 05508-090, Brazil\\
}

\date{Accepted 2019 March 25. Received 2019 February 21; in original form 2018 November 13}

\pubyear{2019}

\begin{document}
\label{firstpage}
\pagerange{\pageref{firstpage}--\pageref{lastpage}}
\maketitle

\begin{abstract}

Globular clusters (GCs) are evolved stellar systems containing entire populations of millisecond pulsars (MSPs), which are efficient $\gamma$-ray emitters. Observations of this emission can be used as a powerful tool to explore the dynamical processes leading to binary system formation in GCs. In this work, 9 years of \textit{Fermi} Large Area Telescope data were used to investigate the $\gamma$-ray emission from all GCs in the Milky Way. 23 clusters were found as $\gamma$-ray bright, with 2 of them never having been reported before. It was also found that magnetic braking probably has a smaller impact on the formation rate of binary systems in metal-rich GCs than previously suggested, while a large value for the two-body encounter rate seems to be a necessary condition. The influence of the encounter rate per formed binary was for the first time explored in conjunction with $\gamma$-ray data, giving evidence that if this quantity is very high, binary systems will get destroyed before having time to evolve into MSPs, thus decreasing the total number of MSPs in a GC. No extended emission was found even for clusters  whose optical extent is $\approx 0.5^{\circ}$; all of them are point-like sources spatially in agreement with the optical cores of the GCs, supporting previous X-rays results of heavier objects sinking into the clusters' cores via dynamical friction. The possibility of extrapolating these results to ultra-compact dwarf galaxies is discussed, as these systems are believed to be the intermediate case between GCs and dwarf galaxies.

\end{abstract}


\section{Introduction}
\label{Intro}

Since the first detection of $\gamma$-rays from 47 Tucan\ae $ $ with the \textit{Fermi} Large Area Telescope (LAT) \citep{abdo2009detection}, globular clusters (GCs) have become a new class of $\gamma$-ray source. Previous observations in the 90's with the Energetic Gamma-ray Experiment (EGRET) aboard the Compton Gamma-ray observatory found no signal of $\gamma$-ray emission from these sources but resulted in important flux upper limits for more than a dozen of them \citep{michelson1994egret}. By studying GCs in $\gamma$-rays, one can learn about the dynamical evolution of these systems, as well as the mechanisms behind the formation of their millisecond pulsar (MSP) populations. 

The $\gamma$-ray emission from GCs is attributed to their large number of MSPs \citep{bednarek2007high,abdo2010population,freire2012pulsar,caraveo2014gamma}, which are known to be efficient $\gamma$-ray emitters \citep{chen1991gamma,harding2005high,caraveo2014gamma}. The recent detection of pulsed $\gamma$-ray emission from some GCs has further strengthened this connection \citep{freire2011fermi,johnson2013broadband}. The populations of MSPs in these systems\footnote{List containing all known pulsars in GCs: \url{https://www.naic.edu/~pfreire/GCpsr.html}}, as the descendants of low-mass X-ray binaries (LMXBs), are believed to be formed as a natural consequence of frequent stellar encounters \citep{pooley2003dynamical}, with the high number density of stars within a GC providing an excellent laboratory to test scenarios for compact binary formation. Evidence favoring this dynamical origin, i.e., positive correlations between the cluster's $\gamma$-luminosity ($L_{\gamma}$) with its stellar encounter-rate ($\Gamma$) and with its metallicity ([Fe/H]) have been found in the past few years \citep{abdo2010population,hui2010dynamical,hui2010fundamental,bahramian2013stellar,hooper2016gamma,lloyd2018gamma}. In interpretations of the former, the stars are assumed to be captured one by the other at a rate proportional to $\Gamma \propto \rho_0^{3/2}r_c^2$, where $\rho_0$ is the central luminosity density and $r_c$ is the cluster core radius \citep{verbunt2003binary}. In the latter, for a metal-rich cluster, magnetic braking can be more efficient, facilitating orbital decay in binary systems as well as a higher probability of MSP formation \citep{hui2010fundamental,tam2016gamma}.

In contrast with other pulsars, MSPs begin as spun-down neutron stars in binary systems where the companion star is massive enough to evolve into a giant and overflow the Roche limit of the system \citep{lorimer2001binary_and_MSP}. The neutron star is then spun-up and wakes up as a recycled pulsar by accreting matter and increasing angular momentum at the expense of the orbital angular momentum of the binary system \citep{alpar1982new}. Due to their weak surface magnetic fields, MSPs lose their larger store of rotational kinetic energy much more slowly than common pulsars \citep{lorimer2005cambridge}, remaining luminous for up to billions of years. $\gamma$-ray photons are mainly produced in the magnetosphere of MSPs, where inverse Compton scattering, synchrotron and curvature radiation are the main physical processes behind this emission \citep{sturrock1971polarcap,harding1978curvature,arons1983slotgap,cheng1986outergap,bednarek2007high}.

This work presents an analysis of all globular clusters in the Milky Way (\citeauthor{harris1996catalog} \citeyear{harris1996catalog} -- 2010 edition), using 9 years of \textit{Fermi} LAT data as an effort for detecting them and characterizing their $\gamma$-ray emission and MSP formation scenarios. The observations and cuts on \textit{Fermi} LAT data are described in Section \ref{Observations}; followed by the light curve analysis and study of correlations between $\gamma$-ray emission, encounter rate and metallicity, in Section \ref{results}. Section \ref{discussion} discusses the influence of magnetic braking on the formation of MSPs and the possibility of extrapolating the results in this work to ultra-compact dwarf galaxies.

\section{Data selection and analysis}
\label{Observations}

The sample analyzed consists of all 157 known GCs in the Milky Way. For 25 of them, $\gamma$-ray emission was previously described in \cite{3fgl}, \cite{hooper2016gamma} and \cite{zhang2016detection}, although four of these sources were not confirmed here (2MASS-GC02, M15, NGC 6342 and Pal6). Each GC treated here was observed with LAT during a 9-year period ranging from August 5th 2008 to August 5th 2017 (MET 239587201 - 523584005). The data was analyzed using \textit{Fermi} Science Tools v10r0p5, \textit{fermipy} python package v0.16.0 \citep{wood2017fermipy} and Pass 8 \citep{atwood2013pass}, which present better energy and angular resolution as well as an increased effective area and energy range than its predecessor Pass 7. 

Following standard procedures \footnote{\textit{Fermi} science tools and fermipy tutorials: \url{https://fermi.gsfc.nasa.gov/ssc/data/analysis/scitools/} and \url{http://fermipy.readthedocs.io/en/latest/quickstart.html}}, data for each source was selected within a $12^{\circ} \times 12^{\circ}$ region-of-interest (ROI), centered on the GCs positions given in the \code{GLOBCLUST} catalog (\citeauthor{harris1996catalog} \citeyear{harris1996catalog} -- 2010 edition), with energies ranging between 100 MeV and 100 GeV divided into 12 logarithmically spaced energy bins. Photons with energies $> 100$ GeV were not considered, as the spectra of MSPs frequently presents an exponential cutoff behavior above only a few GeV \citep{abdo2010first}. Sources included in the \textit{Fermi} LAT Third Source Catalog (3FGL -- \citeauthor{3fgl}, \citeyear{3fgl}) and lying up to $5^{\circ}$ outside the ROIs were taken into account as well as all sources found with the \textit{fermipy} function \textit{find\_sources(sqrt\_ts\_threshold=5.0, min\_separation=0.4)}. The number of new sources found with \textit{find\_sources()} varied substantially depending on the adopted ROI: for ROIs lying close to the Galactic plane, a number of $\sim 20$ new sources was common; for the other ROIs, the number of new sources was typically $\lesssim 10$. In very few cases, a new source was found closer than $0.4^{\circ}$ from the GC position (e.g.: Palomar 6 and Whiting 1). In these situations, a case-by-case approach was performed, manually including a new source to avoid contamination on the GC flux upper limit measurement. Only events belonging to the \textit{Source} class were used (evclass=128 and evtype=3). The filters applied with $gtmktime$ were DATA\_QUAL > 0 and the recommended instrument configuration for science LAT\_CONFIG == 1. A zenith angle cut of $90^{\circ}$ was applied to avoid contamination from the Earth limb. For modeling the Galaxy and the extragalactic background emission, the Galaxy background model gll\_iem\_v06.fits and the isotropic spectral template iso\_P8R2\_SOURCE\_V6\_v06.txt were adopted.

All sources were investigated by means of binned likelihood analysis (\textit{gtlike} tool -- MINUIT algorithm). To quantify the significance among the detections, a test statistic (TS) was calculated, defined as $TS = 2(\mathcal{L}_1 - \mathcal{L}_0)$, where the term inside parentheses is the difference between the maximum log-likelihoods with ($\mathcal{L}_1$) and without ($\mathcal{L}_0$) modeling the source. The chosen criteria for detection was $TS>25$, corresponding formally to a significance slightly above $4\sigma$ \citep{mattox1996likelihood}. If the detected GC belonged to 3FGL, its spectrum was modeled according to its description there. If a detected GC was not included in 3FGL (post-3FGL clusters, from now on), its spectrum was modeled with a power-law. For all sources lying within a radius of $5^{\circ}$ from the center of the ROIs, the normalization parameter was left free to vary.

\begin{table}
\centering
\begin{tabular}{l|c|c|c|c}
Cluster & SM & Luminosity & Energy flux & TS \\
       &  & $10^{34}$erg s$^{-1}$ & $10^{-12}$erg cm$^{-2}$s$^{-1}$ \\
\hline           
47 Tuc  & LP & $ 6.26  \pm  0.19 $ & $ 25.81  \pm  0.77 $ &  5604 \\
Terzan5  & LP & $ 42.39  \pm  1.54 $ & $ 74.41  \pm  2.71 $ &  3854 \\ 
M62  & LP & $ 8.96  \pm  0.55 $ & $ 16.2  \pm  0.99 $ &  1036 \\ 
NGC6388  & LP & $ 29.34  \pm  1.24 $ & $ 25.01  \pm  1.06 $ &  880 \\ 
$\Omega$ Cent & LP & $ 3.57  \pm  0.26 $ & $ 11.03  \pm  0.8 $ &  863 \\ 
2MS-GC01 & LP & $ 8.89  \pm  0.54 $ & $ 57.33  \pm  3.46 $ &  731 \\
NGC6440 & PL & $ 25.64  \pm  1.4 $ & $ 29.66  \pm  1.62 $ &  518 \\ 
NGC6316 & PL & $ 22.38  \pm  1.62 $ & $ 17.29  \pm  1.25 $ &  276 \\ 
NGC6441 & PL & $ 27.37  \pm  1.86 $ & $ 17.0  \pm  1.16 $ &  331 \\ 
NGC6752 & PL & $ 1.02  \pm  0.1 $ & $ 5.32  \pm  0.54 $ &  149 \\ 
NGC6652 & LP & $ 4.19  \pm  0.58 $ & $ 3.5  \pm  0.48 $ &  129 \\ 
M80 & PL & $ 7.10  \pm  0.95 $ & $ 5.94  \pm  0.79 $ &  92 \\ 
NGC2808 & PL & $ 4.89  \pm  0.67 $ & $ 4.43  \pm  0.61 $ &  81 \\ 
NGC6541 & PL & $ 3.34  \pm  0.44 $ & $ 4.97  \pm  0.66 $ &  78 \\ 
NGC6717 & PL & $ 1.76  \pm  0.38 $ & $ 2.92  \pm  0.64 $ &  31 \\ 
\hline
Glimp01 & PL & $ 9.83  \pm  1.29 $ & $ 46.54  \pm  6.09 $ &  286 \\ 
Glimp02 & PL & $ 8.6  \pm  1.07 $ & $ 23.74  \pm  2.96 $ &  98 \\ 
NGC6397 & PL & $ 0.35  \pm  0.05 $ & $ 5.5  \pm  0.79 $ &  64 \\ 
NGC6139 & PL & $ 8.22  \pm  1.31 $ & $ 6.73  \pm  1.07 $ &  59 \\ 
M12 & PL & $ 0.88  \pm  0.17 $ & $ 3.20  \pm  0.61 $ &  43 \\ 
M5 & PL & $ 1.52  \pm  0.33 $ & $ 2.25  \pm  0.49 $ &  31 \\ 
\hline
M14 & PL & $ 3.17  \pm  2.49 $ & $ 3.06  \pm  2.4 $ &  29 \\ 
M79 & PL & $ 3.61  \pm  0.82 $ & $ 1.81  \pm  0.41 $ &  26 \\ 
\hline

\end{tabular}

\caption{ Observations of 23 globular cluster candidates with \textit{Fermi} LAT in an energy range from 100 MeV up to 100 GeV. The upper panel displays the 3FGL associations, the middle panel shows the post-3FGL associations and the bottom panel shows the new candidates for GC. The adopted spectral models (SM) are indicated on the second column, where $PL = Power Law$ and $LP = LogParabola$. The computation of the luminosities was performed assuming isotropic $\gamma$-ray emission and using the distances available in the \code{GLOBCLUST} catalog.}
\label{Tabela1}
\end{table}

Point sources consistent with the optical center of the GCs were added to the models for each GC in the analysed sample and a TS residuals map was constructed for each one of them. These residuals maps were used together with aperture photometry light curves to obtain a clean sample of GCs, where only sources with a clear indication for an isolated $\gamma$-ray point-like emission spatially coincident with the optical position of the cluster and presenting steady light curves were taken into account. All detections are described in Section \ref{results}, while the TS residuals maps and flux upper limits for all non-detected GCs are shown in Appendix \ref{sec.foo}. 

\begin{figure*}

\includegraphics[width=\linewidth]{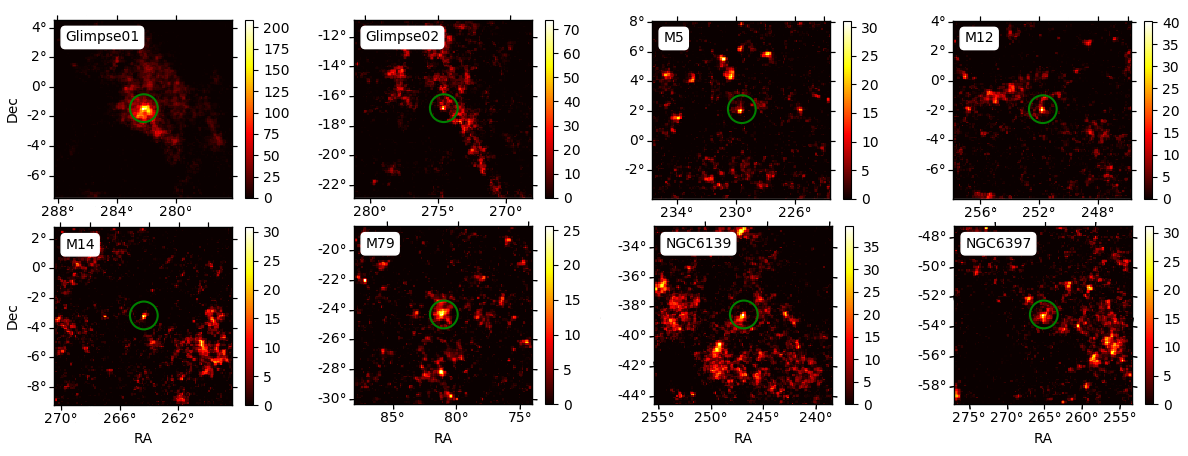}

\caption{Mosaic of TS residuals maps for the firmly detected post-3FGL sources. These maps were generated with \textit{tsmap()} function available in \textit{fermipy}. The green circles guide the readers to the center of the maps.}
\label{TSMosaics}
\end{figure*}

The cases of NGC 6624 and M28 should be considered separately. Both GCs were originally described as $\gamma$-ray bright \citep{abdo2010population,tam2011gamma}. Later, they were observed to host $\gamma$-ray pulsations \citep{freire2011fermi,johnson2013broadband} and were cataloged as individual pulsars in 3FGL. When modeling both clusters, these cataloged pulsars were taken into account, with their normalizations left free to vary. After that, the clusters' emission laid below the detection threshold.

\section{Results}
\label{results}

\subsection{Detections}

Among the 157 GCs, 23 presented $\gamma$-ray emission spatially coincident with the optical center of the clusters. Another 4 clusters (2MASS-GC02, M15, NGC 6342 and Pal6) were previously described as $\gamma$-ray bright in literature, but in this work their significances were found to be below the adopted threshold of TS = 25; the results for these clusters are described in Appendix \ref{sec.foo}. The 23 detected clusters are shown in Table \ref{Tabela1}, which is segmented in three panels: the upper one shows the 15 GCs cataloged in 3FGL; the middle panel shows the post-3FGL clusters, and the last/bottom panel shows the new GCs candidates detected in this work.

\subsubsection{Fermi-LAT 8 years source list}

Cross checking the detections in this work with the ones available on the preliminary LAT 8-year Point Source List (FL8Y\footnote{\url{https://fermi.gsfc.nasa.gov/ssc/data/access/lat/fl8y/}}), some differences are found. Six sources (2MASS-GC02, M92, NGC 362, NGC 6304, NGC 6342 and Terzan 1) associated to GCs in FL8Y are not significant detections in this work; and only one of the GCs detected in this work is not listed in FL8Y (M79). All of these clusters (with exception of Terzan 1) are very close to the detection threshold. The found differences may be related to the different likelihood method adopted in FL8Y (weighted likelihood\footnote{\url{https://fermi.gsfc.nasa.gov/ssc/data/access/lat/fl8y/FL8Y_description_v8.pdf}}) and may vary depending on the analyzed region. The clusters 2MASS-GC02, NGC 362 and Terzan 1, for instance, are located on very complicated regions of the sky with bright diffuse emission. Also M92 and M79 are $\sim 1^{\circ}$ apart from very bright sources. Future analyses including the 4th source catalog of the \textit{Fermi}-LAT (4FGL, in preparation) will benefit from a new Galactic diffuse $\gamma$-ray emission model based on Pass 8 data, allowing for better results.

\subsection{Point-like sources}
\label{point_sources}

The spatial consistency between the $\gamma$-ray and optical/infrared emission for the firmly detected post-3FGL sources can be checked in Figure \ref{TSMosaics}, where the low-energy centers of the GCs are always coincident with the centers of the TS residuals maps. All maps have evidence for $\gamma$-ray emission with a maximum lying less than $0.15^{\circ}$ ($\sim 2$ pixels) from their centers. TS maps for 3FGL clusters are not shown here, as their emission is generally easily visible in their counts maps. The two new sources found in this work, M14 and M79, had their best fitted positions obtained with the function \textit{localize()}, available in \textit{fermipy}, which found a $\gamma$-optical/infrared spatial separation within $1\sigma$ uncertainty radius for both cases (Table \ref{Tabelafindsrc}). 

In order to test the accidental coincidence rate in the analyzed sample, 200 pre-selected points in RA and Dec (blank fields) were randomly chosen in the sky with $|b| > 20^{\circ}$ and being at least $1^{\circ}$ apart from any known 3FGL source. These blank fields were analyzed exactly in the same way as the main GCs analyses and on $4.5\%$ of them a point source with TS above the threshold was detected. This value corresponds to an upper limit on the false positive rate of the associations with GCs shown in Table \ref{Tabela1} and should not be confused with the detection significance of the sources. The 3FGL clusters here were assumed to be truly associated to GCs and are not included in this false alarm association rate.

\begin{table}
\centering
\begin{tabular}{l|c|c|c|c}
Cluster & RA & Dec & $r_{1\sigma}$ & $\alpha$ \\
\hline           
M14 & $264.412^{\circ}$  & $-3.238^{\circ}$ & $0.035^{\circ}$ & $0.014^{\circ}$ \\
M79 & $81.120^{\circ}$ & $-24.410^{\circ}$ & $0.140^{\circ}$ & $0.134^{\circ}$
\end{tabular}
\caption{ Best fit position for M14 and M79. $r_{1\sigma}$ is the $1\sigma$ error circle radius and $\alpha$ is the angular separation between the position of the $\gamma$-ray detection and the optical center of the cluster. In both cases $\alpha$ lies inside the $1\sigma$ uncertainty region. Positions are given in J2000 coordinates.}
\label{Tabelafindsrc}
\end{table}

\subsection{Extended source analysis}

For $\Omega$ Centauri, 47 Tucan\ae $ $ and NGC 6397, the three $\gamma$-ray bright GCs with largest optical angular diameter ($\gtrsim 0.5^{\circ}$), extended emission models were tested. The extended emission templates were created in two ways: with DSS optical maps available in NASA's SkyView\footnote{\url{https://skyview.gsfc.nasa.gov/current/cgi/query.pl}} virtual telescope \citep{mcglynn1998skyview} and with 25 2D-Gaussian source templates with sizes ranging from $0.003^{\circ}$ to $1^{\circ}$. For all tests, the likelihood ratio method favored the point-like model instead of the extended emission models, which, in the most optimistic cases, presented TS values of only $TS_{\rm ext} = 0.73$ for $\Omega$ Centauri ($R_{68} = 0.055 \pm 0.031$), $TS_{\rm ext} = 0.38$ for 47 Tucan\ae $ $ ($R_{68} = 0.033 \pm 0.022$) and $TS_{\rm ext} = 0.00$ for NGC 6397 ($R_{68} = 0.003 \pm 0.080$). These results are in agreement with heavier objects sinking into the clusters' cores via dynamical friction \citep{fregeau2003monte}, as both binaries and single MSPs are significantly more massive than typical stars in a GC. The same conclusion is reached from X-ray observations of 47 Tucan\ae $ $ \citep{edmonds2003extensive,heinke2005deep}, where the cluster's X-ray source population is highly concentrated in its core.

\subsection{Light curves}
\label{globclustcharac}

\begin{figure*}
\centering
\includegraphics[width=\linewidth]{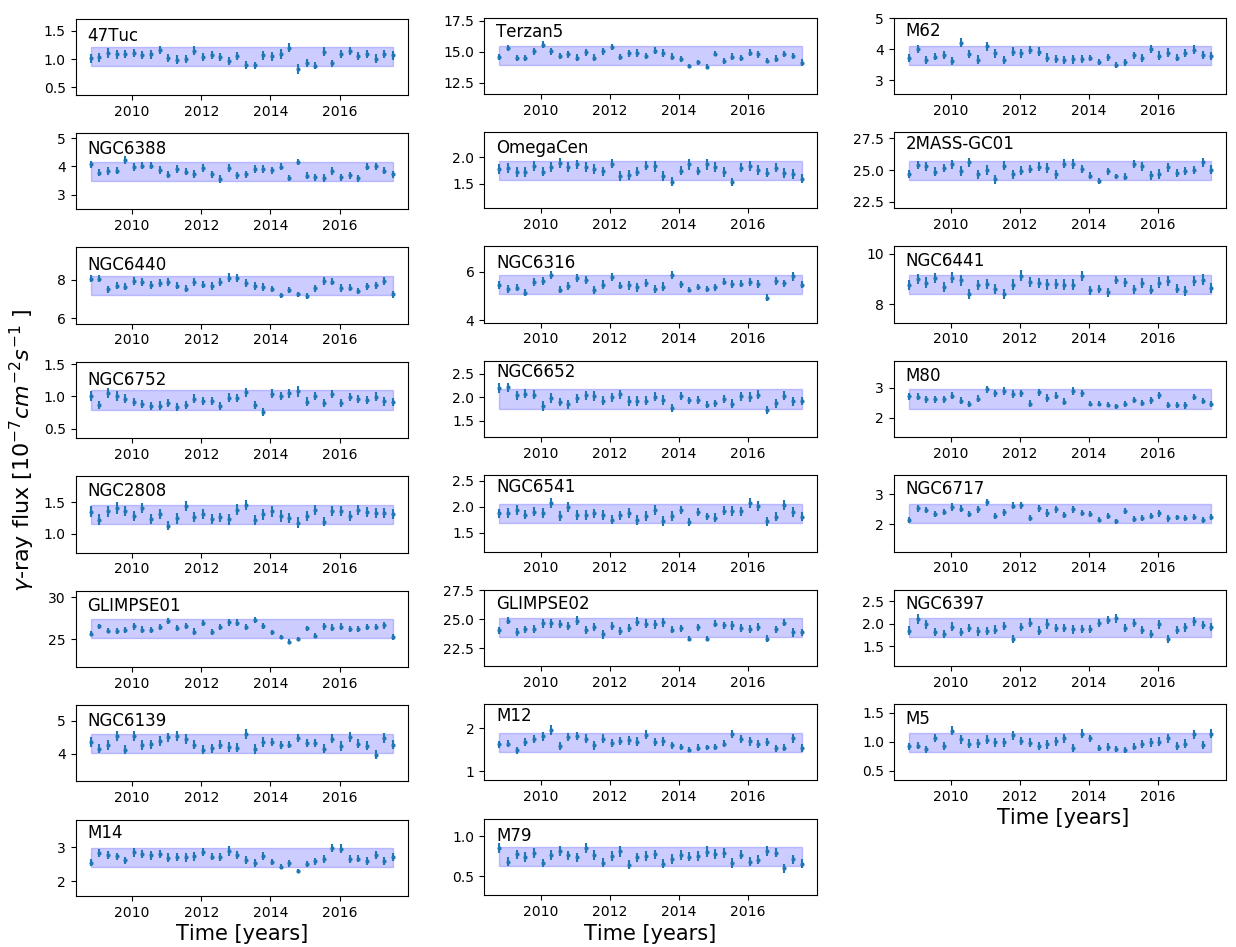}
\caption{Aperture photometry light curves for the 23 $\gamma$-ray bright GC. Each bin corresponds to 3 months of data in an energy range from 100 MeV to 100 GeV and aperture radius of $1^{\circ}$. All sources exhibit weak variability, showing no flares with peaks significantly above the $2\sigma$ deviation level (blue band). None of the clusters were excluded from the analyzed sample based on their light curves.}
\label{Ap_Phot_Lightcurves}
\end{figure*}

The intrinsic variability of MSPs completely disappears when the observations span through timescales much longer than the MSP typical revolution time. A light curve with bins of months is expected to be quiescent \citep{abdo2010population}, where its variability is attributed only to statistical fluctuations. High levels of variability in a light curve with such timescales are unlikely to be associated with a GC; they could nevertheless be due to a background active galactic nucleus.

To test for variability, aperture photometry light curves were created with \textit{gtbin} for every single $\gamma$-ray GC. The results are shown in Figure \ref{Ap_Phot_Lightcurves}, where the data was binned in 3-month intervals, the chosen aperture radius was $1^{\circ}$ and the spectral index was kept fixed at 2. All light curves analyzed presented quiescent behavior, as expected for GCs, with all data points lying within the $2\sigma$ standard deviation level (blue band).

\subsection{Spectral emission models}
\label{spectra}

The high energy $\gamma$-ray spectra for 19 of the 23 detected GCs are shown in Figure \ref{fig:SEDs} in Appendix \ref{sec.foo}. All spectra are reasonably well fitted by a Logparabola or a power-law model, in agreement with the GCs discussed in \cite{abdo2010population}. As no significant deviation from such models is observed, the traditional interpretation that the $\gamma$-ray emission observed is coming from populations of MSPs was assumed throughout this work. Other models are discussed below.

One possible mechanism for producing $\gamma$-rays in GCs is inverse Compton scattering (ICS) by a relativistic population of electrons in the intracluster medium \citep{bednarek2007high}. The spectra derived from such model, however, predict a hardening of the spectrum around 1-10 GeV \citep{bednarek2007high,lloyd2018gamma}. In this work, no strong evidence for a spectral hardening was found in this band (Figure \ref{fig:SEDs}). 

Although dark matter annihilation has been proposed as another source of $\gamma$-rays in GCs \citep{brown2018understanding}, the lack of strong dynamical evidence \citep{moore1996constraints,hacsegan2005acs} suggests such processes cannot explain the majority of the $\gamma$-ray emission.

The results shown in Figure \ref{Ap_Phot_Lightcurves} reduces the possibility of associating the observed $\gamma$-ray emission with cataclysmic variables within the clusters, as such sources are transient in nature \citep{abdo2010gammaNova} and no significant variability was observed in the analyzed light curves.

\begin{figure*}
\includegraphics[width=\linewidth]{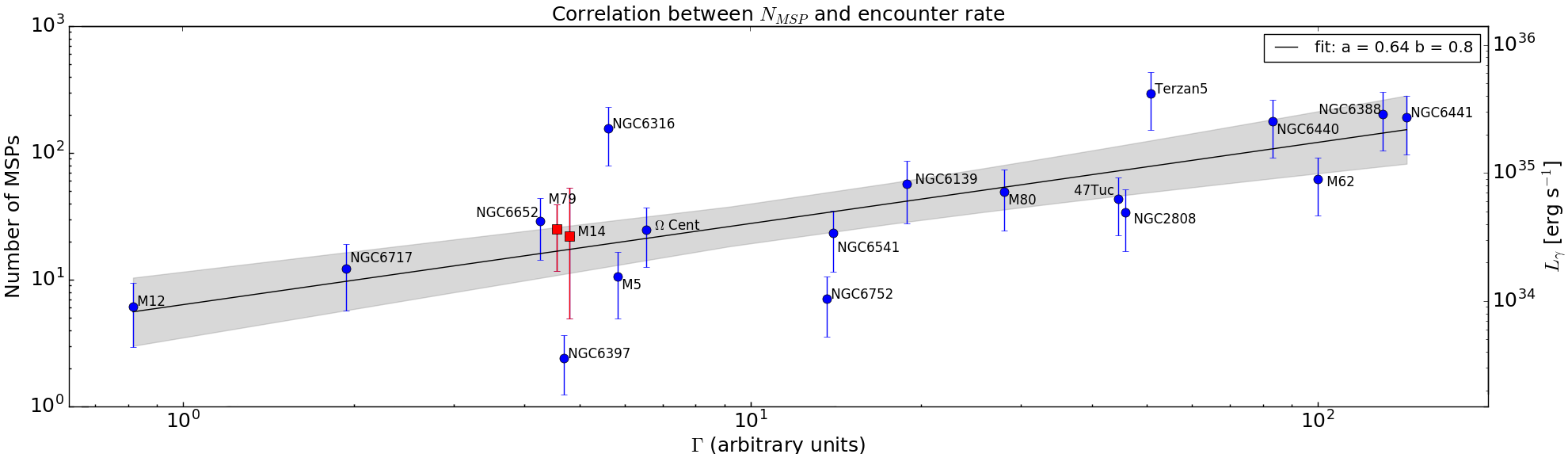}

\vspace{0.5cm}

\includegraphics[width=\linewidth]{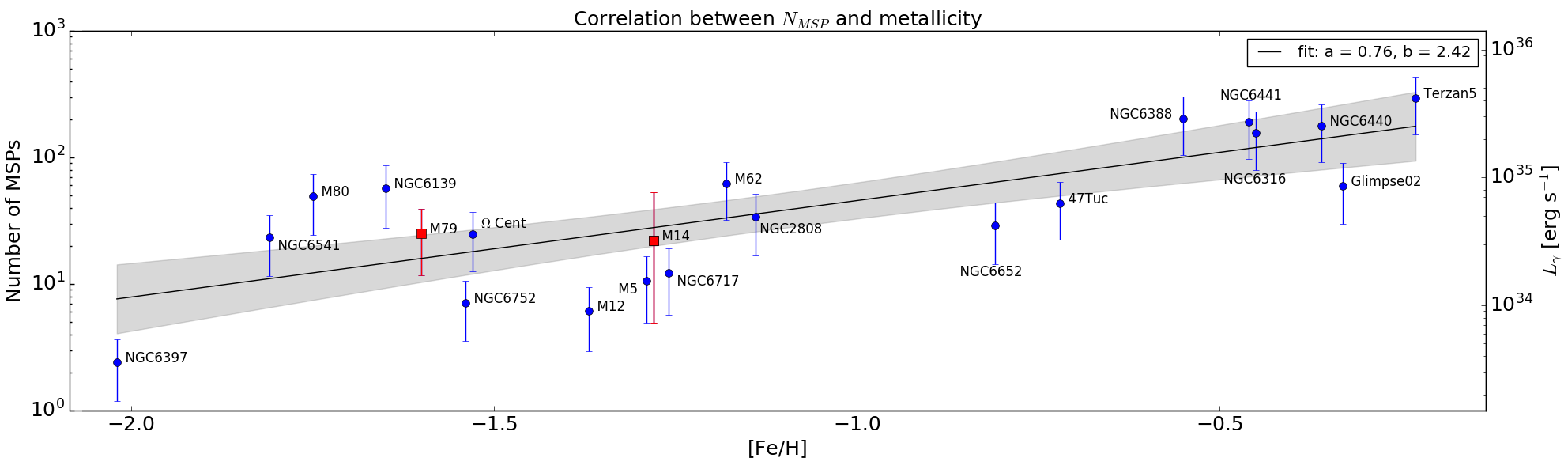}

\caption{Correlation plots for $N_{\rm MSP} \times \Gamma$ and $N_{\rm MSP} \times [Fe/H]$ for the $\gamma$-ray bright GCs described in this work. The two new candidates for GCs found in this work are plotted as red squares. Sources lacking values of metallicity, central luminosity density or cluster's core radius in \code{GLOBCLUST} catalog were neglected. The shaded areas are the $1\sigma$ confidence bands. For a version of the upper panel plot including upper limits, see Appendix \ref{sec.foo}.}
\label{PL_x_line}
\end{figure*}

\subsection{Correlated quantities}
\label{correlatedQuantities}

Parameters such as the two-body encounter rate $\Gamma$ and metallicity [Fe/H] are expected to influence the formation rate -- and so the total number -- of MSPs in a GC (see Section \ref{Intro}). For estimating the total number of MSPs, $N_{\rm MSP}$, within a cluster, a simple calculation was performed \citep{abdo2010population}: 
\begin{equation}
N_{\rm MSP} = \frac{L_{\gamma}}{\langle\dot{E}\rangle \langle\eta_{\gamma}\rangle},
\label{eq:1}
\end{equation}
where $L_{\gamma}$ is the cluster's isotropic $\gamma$-ray luminosity, $\langle \dot{E}\rangle$ is the average power emitted during the spin down of MSPs and $\langle \eta_{\gamma} \rangle$ is the average efficiency with which the spin down power is converted into $\gamma$-ray luminosity. The isotropic energy luminosity was simply calculated as $L_{\gamma} = 4\pi r^2 \epsilon_{\gamma}$, where $\epsilon_{\gamma}$ is the measured energy flux (Table \ref{Tabela1}) and $r$ is the distance to the cluster taken from \citeauthor{harris1996catalog} (\citeyear{harris1996catalog}; 2010 edition). The average spin-down power and average spin-down-to-$\gamma$-ray efficiency were adopted as $\langle\dot{E}\rangle = (1.8\pm 0.7)\times 10^{34}$ erg s$^{-1}$ and $\langle \eta_{\gamma} \rangle = 0.08$ for all clusters \citep{abdo2010population}. These values were estimated from comparisons of the $\log \dot{P}$ distributions of Galactic field MSPs with the accelerated corrected $\log \dot{P}$ distribution for MSPs in 47 Tucan\ae $ $ and by the average $\eta_{\gamma}$ efficiency of the nearest MSPs to date as described in \cite{abdo2009detection}. Equation (\ref{eq:1}) is a rough estimate of $L_{\gamma}$, as in some cases $L_{\gamma}$ may be dominated by the emission of a single MSP \citep{freire2011fermi,tam2011gamma,wu2013search}. Nevertheless, it should be good enough to at least establish an upper limit for the actual number of MSPs in a GC.

The scatter plots for $N_{\rm MSP}$ versus $\Gamma$ and $N_{\rm MSP}$ versus [Fe/H] are shown in Figure \ref{PL_x_line}. For these plots, all $\gamma$-ray bright GCs with data values for metallicity and central luminosity density (needed for calculating $\Gamma$) available in the \code{GLOBCLUST} catalog were used. Note that the encounter rate is in arbitrary units and was normalized such that $\Gamma = 100$ for M62, as done in \cite{abdo2010population}.

A simple linear least-squares regression was performed to the data displayed in Fig. \ref{PL_x_line}. In the upper panel, 
\begin{equation}
\log N_{\rm MSP} = a\log\Gamma + b
\end{equation}
where $a=0.64 \pm 0.15$, $b = 0.80 \pm 0.20$ with a mean deviation of the data about the model of $\Delta (\log N_{\rm MSP}) = 0.40$ dex and a Pearson correlation coefficient $P_{\rm corr}$ and $p$-value for testing non-correlation $p_{\rm n-c}$ of $P_{\rm corr} = 0.72$ and $p_{\rm n-c} = 0.00034$, respectively. In the bottom panel, 
\begin{equation}
\log N_{\rm MSP} = a[Fe/H] + b
\end{equation}
with $a =  0.76 \pm 0.14$, $b = 2.42 \pm 0.17$, $\Delta (\log N_{\rm MSP}) = 0.37$ dex, $P_{\rm corr} = 0.76$ and $p_{\rm n-c} = 0.000062$. Both correlations indicate that the $\gamma$-ray luminosity (or $N_{\rm MSP}$, since $N_{\rm MSP} \propto L_{\gamma}$) of a cluster increases with its $\Gamma$ and/or metallicity.

\begin{figure*}
\includegraphics[width=\linewidth]{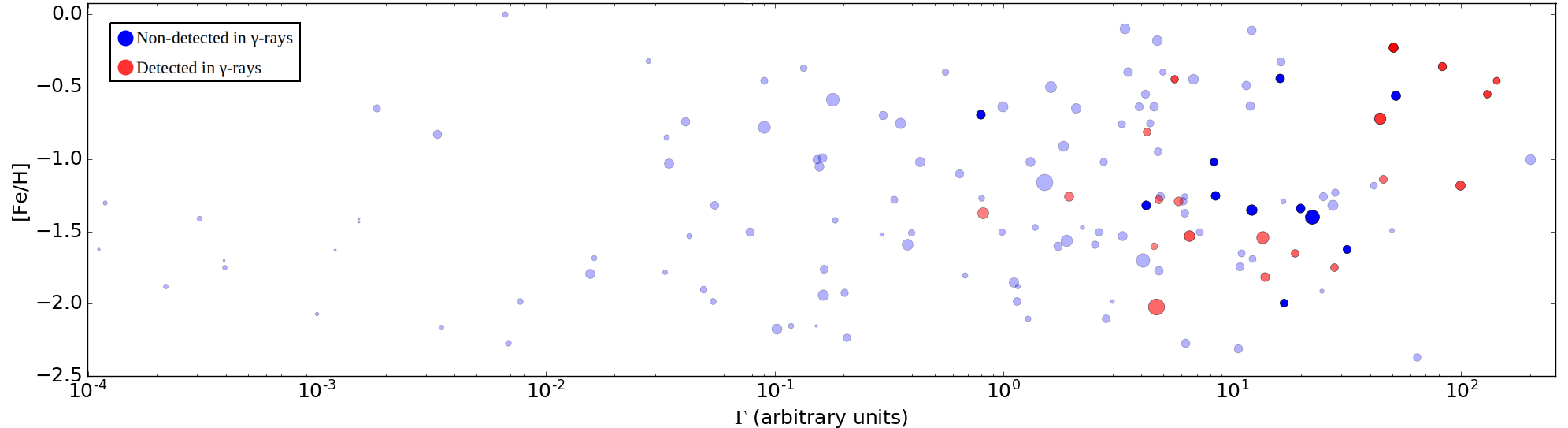}
\caption{ Influence of [Fe/H] and $\Gamma$ for 143 GCs in the Milky Way (those for which metallicity and encounter rate can be taken or calculated from data available in \code{GLOBCLUST}). The size of each point scales with distance: the closer the cluster, the bigger the point. \textbf{Blue points}: GCs non-detected by \textit{Fermi}-LAT. Clusters with $\Lambda > 350$ are represented in dark blue. These sources are expected to be faint in $\gamma$-rays (see section \ref{BinaryDisr}). \textbf{Red points}: GCs detected in $\gamma$-rays. In this case, the darker the point, the bigger the $\gamma$-ray energy flux. Detections are concentrated towards high values of encounter rate.}
\label{Predictions_Encounter}
\end{figure*}

\begin{figure}
\centering
\includegraphics[width=\linewidth]{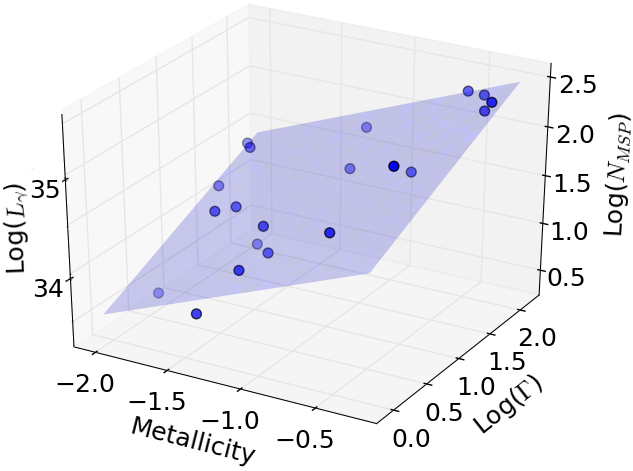}

\vspace{0.2cm}

\includegraphics[width=\linewidth]{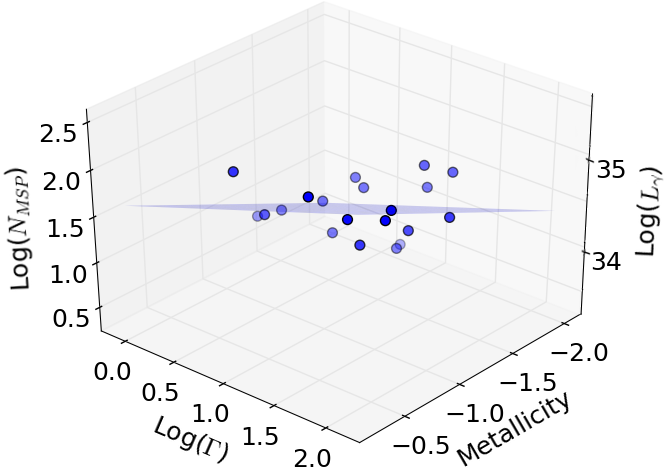}

\caption{ Fit of the plane $\log N_{\rm MSP} = 0.60\times[Fe/H] + 0.39\times\log\Gamma + 1.78$ viewed from different angles.}
\label{plano}
\end{figure}

To test if these correlations are valid for all GCs in the Milky Way, Figure \ref{Predictions_Encounter} shows the $[Fe/H] \times \Gamma$ scatter plot. As can be seen, all detected $\gamma$-ray clusters have a relatively large encounter rate, while their metallicities assume very dispersed values. This result may indicate that the magnetic braking effect is not significantly enhanced in metal-rich clusters, having a smaller role in the compact binary system formation rate than previously suggested \citep{hui2010fundamental,tam2016gamma}. Although the metallicity and the $N_{MSP}$ (or $L_{\gamma}$) indeed present a correlation, as shown in Figures \ref{PL_x_line} and \ref{plano}, the high metallicity may not be interpreted as causing the formation of MSPs (via magnetic braking). If the metallicity increased the MSP formation, then the detections (red dots) in Figure \ref{Predictions_Encounter} should be concentrated in the upper part of the plot. On the other hand, the presence of MSPs is an indicative of a past full of supernovae explosions, which could enhance the clusters' environments with metals (see section \ref{discussion} for a discussion). 

A correlation between $N_{\rm MSP}$, $\Gamma$ and [Fe/H] in a plane was also explored in a three-dimensional log space. Figure \ref{plano} shows the fit from different angles. For testing its goodness of fit, a reduced chi-squared coefficient was calculated, giving $\chi^2_{\nu} = 2.67$. The plane is described by 
\begin{equation}
\log N_{\rm MSP} = a[Fe/H] + b\log \Gamma + c
\end{equation}
where $a = 0.60 \pm 0.14$, $b = 0.39 \pm 0.12$ and $c = 1.78 \pm 0.27$, with a mean deviation of the data about the model of only $\Delta (\log N_{\rm MSP}) = 0.29$ dex. In comparison with the fits in Figure \ref{PL_x_line}, the scatter here is smaller, suggesting that a plane is a better description of the data. The advantage of displaying the data in this way is that once two low-energy observables are obtained ($\Gamma$ and $[Fe/H]$), one can roughly constrain the $\gamma$-luminosity of a GC.

Planes relating $L_{\gamma}$ with $\Gamma$ and $u_{\rm sp}$ (the soft photon energy density), and with $[Fe/H]$ and $u_{\rm sp}$, were proposed in the past \citep{hui2010fundamental}, but the data used there was limited to less than 2 years of \textit{Fermi}-LAT observations, resulting in high dispersion plots.

\subsection{Secondary encounters and binary disruption}
\label{BinaryDisr}

Besides the observed correlation between $N_{\rm MSP}$ and $\Gamma$, there are some aspects of the formation and evolution of MSPs not described by $\Gamma$, like the effects caused by secondary encounters. Once the binary is formed, it may undergo subsequent encounters, which may disrupt the system or even exchange binary members. 

\begin{figure}
\centering
\includegraphics[width=\linewidth]{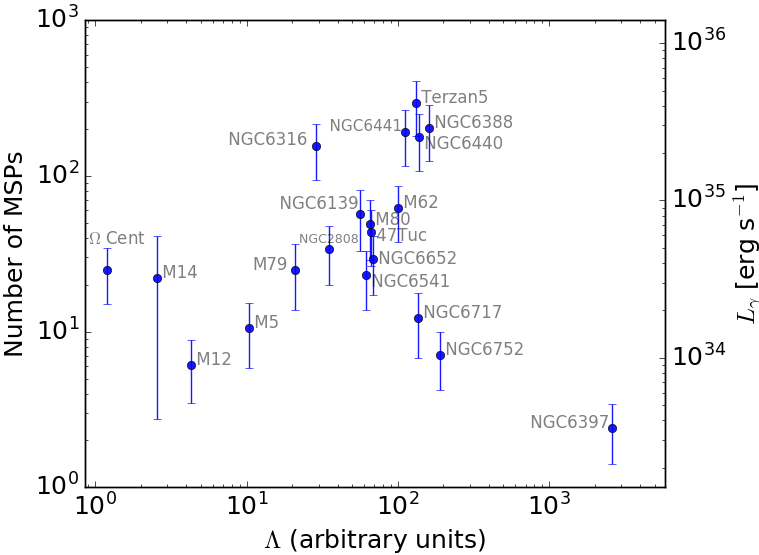}
\caption{ Impact of $\Lambda$ on the populations of MSPs. The largest populations of MSPs are concentrated in clusters with intermediate values of $\Lambda$, indicating that MSPs are preferentially formed in secondary exchange encounters.}
\label{EncRpb}
\end{figure}

Analogously to the estimation of $\Gamma$ described in section \ref{Intro}, one can estimate the encounter rate per formed binary as $\Lambda \propto \sqrt{\rho_0}/r_c$ \citep{verbunt2003new,verbunt2014disruption}, where $\rho_0$ and $r_c$ are the central luminosity density and core radius respectively. In GCs with large values for $\Lambda$, the lifetime of binaries should be relatively short ($\tau = 1/\Lambda$) before being disrupted or undergoing an exchange. The evolution of LMXBs in such clusters may be interrupted before their neutron stars become completely recycled, which may affect their overall population of MSPs. This behavior is perhaps what is seen in Figure \ref{EncRpb}, where a drop in $N_{\rm MSP}$ is evident for very large values of $\Lambda$ (normalized such that $\Lambda = 100$ for M62). The high $\Lambda$ values for NGC 6752 and NGC 6397 can also explain why these clusters have such low $\gamma$-ray luminosity and appear as outliers in Figure \ref{PL_x_line}.

Interestingly, GCs with intermediate values for $\Lambda$ are those with the largest populations of MSPs (Figure \ref{EncRpb}). This suggests that MSPs are preferentially formed in secondary exchange encounters, as only clusters with intermediate or high $\Lambda$ are likely to host pulsar binaries formed in this way \citep{verbunt2014disruption}.

High values for $\Lambda$ may also be the explanation for many of the non-detections shown in Figure \ref{Predictions_Encounter}. Many of the clusters with high $\Gamma$, which are roughly expected to be $\gamma$-ray bright, also have high $\Lambda$, which may negatively affect their MSP population. Non-detected clusters with $\Lambda > 350$ are plotted in dark blue in Figure \ref{Predictions_Encounter}.

\section{Discussion}
\label{discussion}

\subsection{Connection between GCs and UCDs}

Ultra-compact dwarf galaxies (UCDs) are a class of stellar system much brighter and more massive than typical globular clusters, but slightly more extended \citep{Drinkwater2003UCD}. Similarities between both classes of objects are extensively discussed in literature \citep{mieske2002ultra,drinkwater2004ultra,forbes2008uniting}, where UCDs are treated as a link between GCs and dwarf galaxies, with the possible presence of dark matter in UCDs being the main difference between them \citep{hacsegan2005acs}. Taking advantage of these similarities, a tentative estimate of the $\gamma$-ray emission for a couple of UCDs was performed. 

To have an idea of their $\gamma$-ray luminosities, some of the densest known UCDs were chosen: M59-UCD3 \citep{liu2015most} and M85-HCC1 \citep{sandoval2015hiding} and the $N_{\rm MSP} \times \Gamma$ correlation discussed in section \ref{results} was applied. Using their central luminosity densities and core radii (the core radius $r_{\rm core}$ was assumed here as $1/5 \times r_{\rm half}$, the half-light radius, which is very close to the mean value for the ratio $r_{\rm core}/r_{\rm half}$ in \code{GLOBCLUST} catalog) given by \cite{sandoval2015hiding}, their values for $\Gamma$ and then $L_{\gamma}$ were estimated (Table \ref{Table3_UCD}). Despite M85-HCC1 presenting a value for $\Gamma$ beyond the fitted range of Figure \ref{PL_x_line}, it was assumed that the correlation was still valid at least as a first approximation. To estimate their energy fluxes, isotropic emission was assumed and the distances to M59 and M85 were taken from \cite{blakeslee2009acs}. Despite their high luminosity, these compact galaxies are so far away that their energy fluxes are extremely low, not detectable by the Fermi-LAT.

\begin{table}
\centering
\begin{tabular}{l|c|c|c|c}

UCD & Luminosity & Energy flux           &  $\Gamma$ & Distance \\
    &erg s$^{-1}$& erg cm$^{-2}$s$^{-1}$ &           &   Mpc    \\
\hline
M59-UCD3 & 1.8$\times 10^{35}$ & 6.7$\times 10^{-18}$ &  123  & $\sim 15$ \\
M85-HCC1 & 13$\times 10^{35}$ & 34$\times 10^{-18}$ &  2848  & $\sim 18$

\end{tabular}
\caption{UCD $\gamma$-ray energy fluxes and luminosities estimated by the encounter rate correlation described in the text. Fluxes are substantially below \textit{Fermi}-LAT threshold.}
\label{Table3_UCD}
\end{table}

These estimations, although naive, may be important in the future when looking for very faint signals as, for example, dark matter annihilation lines in galaxy clusters crowded with UCDs, like Virgo \citep{jones2006discovery} and Abell 1689 \citep{mieske2004ultracompact}, or even in individual dwarf elliptical galaxies. Searches like these were performed in recent years \citep{ackermann2010constraints,ando2012fermi,ackermann2015searching,ackermann2015search} where interesting upper limits for dark matter annihilation models were found.

\subsection{The driving mechanism behind MSP formation}

In a scenario where the magnetic braking effect is significantly enhanced by a metal-rich environment, high $\gamma$-ray fluxes should be seen for clusters with high metallicities; but this is not observed in Figure \ref{Predictions_Encounter}, which shows that metal-rich clusters are not necessarily efficient $\gamma$-ray emitters. This result suggests that a higher metallicity does not imply a significantly larger magnetic braking effect (and thus a higher MSP formation rate). Hypotheses for explaining the high metallicities in clusters crowded with MSP may be related to the feedback of MSPs or their progenitors within the clusters environment.

At least 5 of the 23 clusters described in this work ($\Omega$ Cen, NGC 2808, NGC 6397, NGC 6752 and M5) are abundant in Calcium \citep{lee2009enrichment}. Calcium and other heavy elements can only be supplied to these systems via supernovae explosions \citep{timmes1995galacti}. As the gravitational potential well in present day clusters cannot retain most of the ejecta from such explosions \citep{baumgardt2008influence}, it has been suggested by \cite{lee2009enrichment} that these GCs are most likely relics of what were once the nuclei of primordial dwarf galaxies accreted and disrupted by the Milky Way \citep{bica2006globular}. This supernova enrichment hypothesis is also supported by some GCs formation models, where stellar winds and supernova ejecta within proto-GCs are decelerated to speeds below the clusters' escape velocity by the pressure of the surrounding hot gas in which they are embedded \citep{brown1991formation}. The metallicity of a cluster, in this context, is simply a function of its total number of supernovae (and not necessarily $N_{\rm MSP}$).

\section{Conclusions}

Nine years of \textit{Fermi} LAT data were analyzed, revealing GC candidates characterized by quiescent $\gamma$-ray emission spatially coincident with the optical centers of the clusters. The novelty of this work is mainly:

\begin{itemize}
\item Evidence that metallicity does not have a significant impact on the MSPs formation. If a metal-rich environment was one of the causes of MSP formation (enhancing the magnetic braking), a concentration of detections (red dots) should be seen in the upper part of Figure \ref{Predictions_Encounter}. The results indicate that the MSP formation may be dominated by the encounter rate and encounter rate per binary rather than by an enhanced magnetic braking effect.

\item It was the first time that the encounter rate per formed binary ($\Lambda$) was analyzed in conjunction with the $\gamma$-ray luminosity (Figure \ref{EncRpb}). The resulting insight was that if $\Lambda$ is very high, binary systems will get destroyed before having time to evolve into a MSP, so impacting the total number of MSPs in a GC. 

\item The characterization of a clean sample of 23 $\gamma$-ray bright GC, where 2 of them (M14 and M79) have never been reported before, as well as upper limits on energy flux for all remaining GCs in the Milky Way.

\item No detected cluster presented extended emission; all of them are point-like sources spatially in agreement with the optical core of the GCs. It was confirmed (for the first time in $\gamma$-rays) the X-rays results of heavier objects sinking into the clusters' cores via dynamical friction.

\end{itemize}

\section*{Acknowledgements}

We acknowledge useful discussions with Paulo C. C. Freire, Matthew Kerr, Philippe Bruel, Heitor Ernandes, Ana Chies Santos and Beatriz Barbuy. We thank the anonymous referee for constructive comments which helped to improve the manuscript. This work was supported by CNPq and FAPESP (Funda\c{c}\~ao de Amparo \`a Pesquisa do Estado de S\~ao Paulo, grants 2016/25484-9 and 2017/01461-2).

The \textit{Fermi} LAT Collaboration acknowledges generous ongoing support
from a number of agencies and institutes that have supported both the
development and the operation of the LAT as well as scientific data analysis.
These include the National Aeronautics and Space Administration and the
Department of Energy in the United States, the Commissariat \`a l'Energie Atomique
and the Centre National de la Recherche Scientifique / Institut National de Physique
Nucl\'eaire et de Physique des Particules in France, the Agenzia Spaziale Italiana
and the Istituto Nazionale di Fisica Nucleare in Italy, the Ministry of Education,
Culture, Sports, Science and Technology (MEXT), High Energy Accelerator Research
Organization (KEK) and Japan Aerospace Exploration Agency (JAXA) in Japan, and
the K.~A.~Wallenberg Foundation, the Swedish Research Council and the
Swedish National Space Board in Sweden.
 
Additional support for science analysis during the operations phase is gratefully
acknowledged from the Istituto Nazionale di Astrofisica in Italy and the Centre
National d'\'Etudes Spatiales in France. This work performed in part under DOE
Contract DE-AC02-76SF00515.

\bibliographystyle{mnras}
\bibliography{refs}

\appendix\section{Upper limits and TS residuals maps for non-detected GC}\label{sec.foo}

\begin{figure*}
\centering
\includegraphics[width=\linewidth]{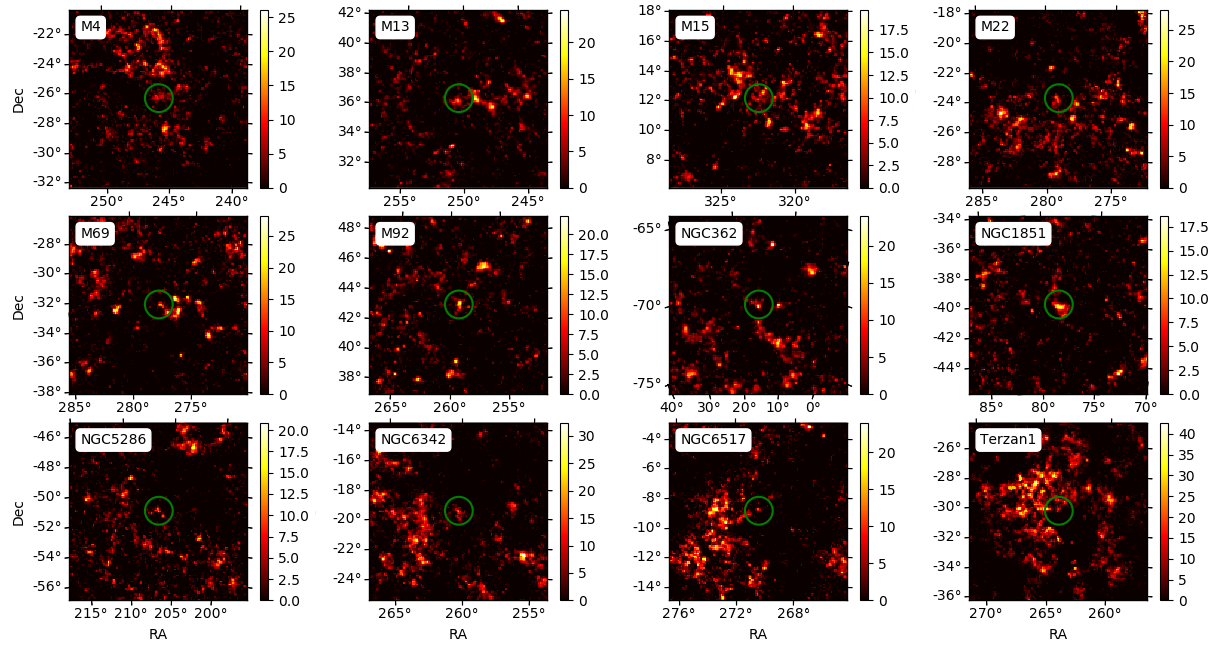}
\caption{TS residuals maps for sources that will be probably associated to GCs within a few years. Their emission were modeled by power-laws and presented significances below $4\sigma$. All maps are centered on the GCs positions given by the GLOBCUST catalog. The green circles guide the readers to the center of the maps.}
\label{future_targets}
\end{figure*}

TS residuals maps for all non-detected GCs are presented below, where the chosen test source was point-like, with an index 2 power-law spectrum. Maps presenting an isolated emission coincident with the optical position of a GC with $6 < TS < 25$ are shown in Figure \ref{future_targets}. These weak signals will probably be associated with GCs with a reasonable significance ($TS>25$) within the next few years. All other non-detected GCs in the Milky Way presented $TS < 6$ or are surrounded by several point-sources with similar significance, making them hard to distinguish. These clusters are shown in Figure \ref{AllGC}. Energy flux upper limits with 95\% confidence levels and integrated over the whole analysis energy range are provided in Tables \ref{Tabela2_1} and \ref{Tabela2_2}, where the sources were assumed to have a power-law spectrum with a spectral index fixed at 2. A few of the sources described below are associated with globular clusters very close to the detection threshold in the preliminary LAT 8-year point source list \footnote{\url{https://fermi.gsfc.nasa.gov/ssc/data/access/lat/fl8y/}} (FL8Y) and will likely be part of the 4th Fermi-LAT catalog (4FGL, in preparation). The reason why these sources have TS slightly above the threshold in FL8Y is likely related to the different likelihood method used for creating the catalog (weighted likelihood).

The $\gamma$-ray spectra for 19 GCs, as discussed in Section \ref{spectra}, are shown in Figure \ref{fig:SEDs}. The spectra of only 4 of the 23 detected clusters were not included, all of them due to problems with low statistics or difficult sky positions. The clusters detected with high significance in $\gamma$-rays are mainly dominated by a logparabola spectral shape, while for the low-significance ones, the logparabola model is not statistically preferred over a power-law. The spectra of 2MASS-GC01, Glimpse 01 and Glimpse 02 may have significant contamination from the Galactic diffuse emission, as these GCs are localized very close to the Galactic plane. The 19 spectra were obtained using the \textit{fermipy} function \textit{sed()}\footnote{\url{https://fermipy.readthedocs.io/en/latest/advanced/sed.html}}, which computes the $\gamma$-ray spectra by performing independent fits for the flux normalization of a source in logarithmic spaced bins of energy (from 100 MeV up to 100 GeV in this case).

In section \ref{correlatedQuantities} a linear regression was performed taking into account only the measurements of $\Gamma$ and $N_{\rm MSP}$ (cf. Fig. \ref{PL_x_line}a). In order to quantify the impact of the non-detections on the results, the Python port of the \code{LINMIX\_ERR} package\footnote{\url{https://github.com/jmeyers314/linmix}} was used. \code{LINMIX\_ERR} is a Bayesian linear regression method that takes into account both measurement errors and non-detections \citep{Kelly07}. The fit incorporating upper limits is given by
\begin{equation}
\log N_{\rm MSP} = (2.17 \pm 0.55) \log \Gamma +(-2.86 \pm 1.07),
\end{equation}
which should be compared with the fit incorporating only measured values (cf. section \ref{correlatedQuantities}),
\begin{equation}
\log N_{\rm MSP} = (0.64 \pm 0.15) \log \Gamma +(0.80 \pm 0.20).
\end{equation}

\begin{figure}
\centering
\includegraphics[width=\linewidth]{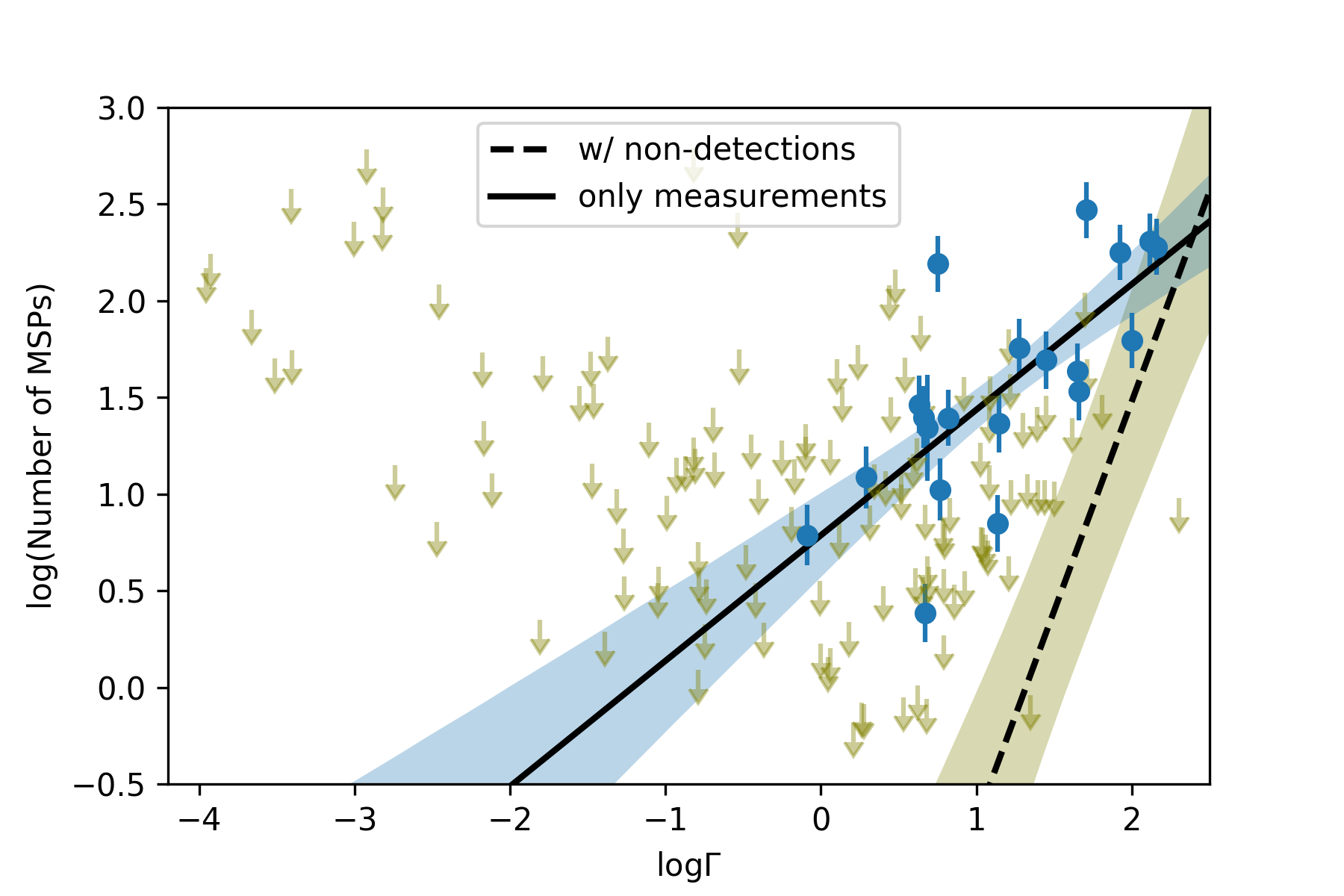}
\caption{Linear regression for the number of MSPs and encounter rates. The solid line shows the fit to the data incorporating only actual measurements, while the dashed line takes into account measurements and non-detections (upper limits). The shaded regions around each line display the $1\sigma$ credibility bands for each fit. }
\label{censored}
\end{figure}

Given that non-detections outnumber the detections by a factor of $\sim 7$, it is not surprising that the two fits are considerably different from each other, as can be seen in Figure \ref{censored}. The fit involving only the detections is in good agreement with what is observed in X-rays and $\gamma$-rays for LMXBs and MSPs in previous works \citep{pooley2003dynamical,abdo2010population,hooper2016gamma}, while the fit incorporating upper limits seems to underestimate the results in these works, giving a much lower number of MSPs per GC. This may be related to the high values of $\Lambda$ found for many of these non-detections, which may diminish the expected $\gamma$-ray flux even if the source has a large value for $\Gamma$ (see section \ref{BinaryDisr} for details).

\begin{figure*}
\centering
\includegraphics[width=\linewidth]{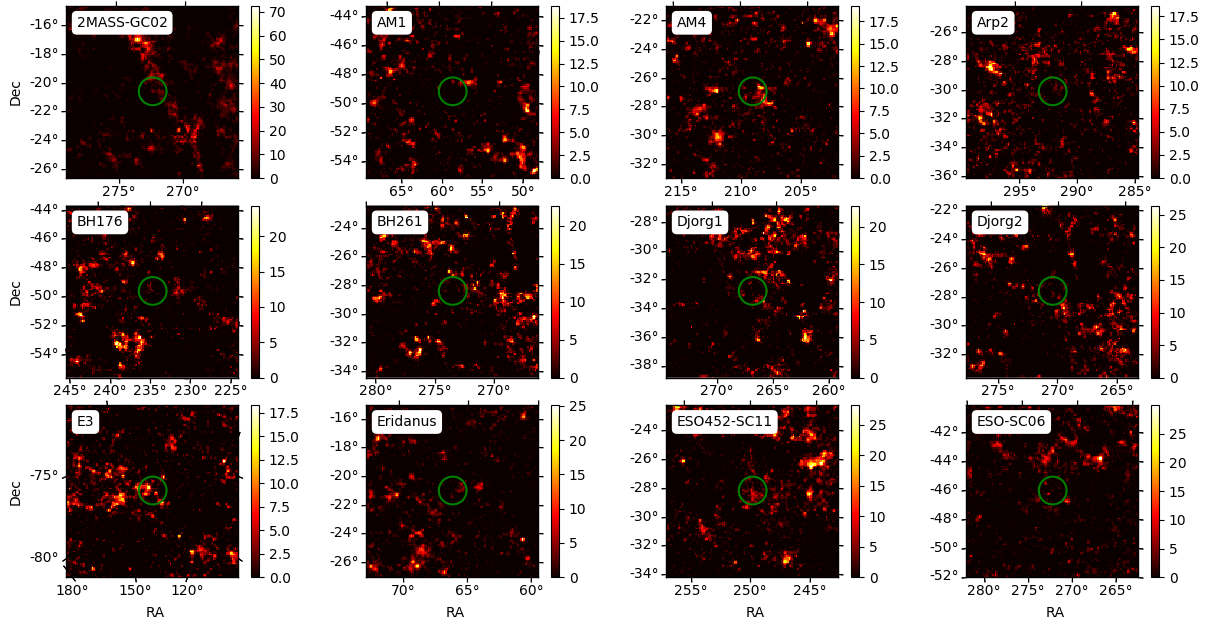}
\caption{TS residuals maps for all Milky Way GCs non-detected in $\gamma$-rays. All maps are centered on the GC positions given by the GLOBCUST catalog. The green circles guide the readers to the center of the maps.}
\label{AllGC}
\end{figure*}

\begin{figure*}
\ContinuedFloat 
\centering
\includegraphics[width=\linewidth]{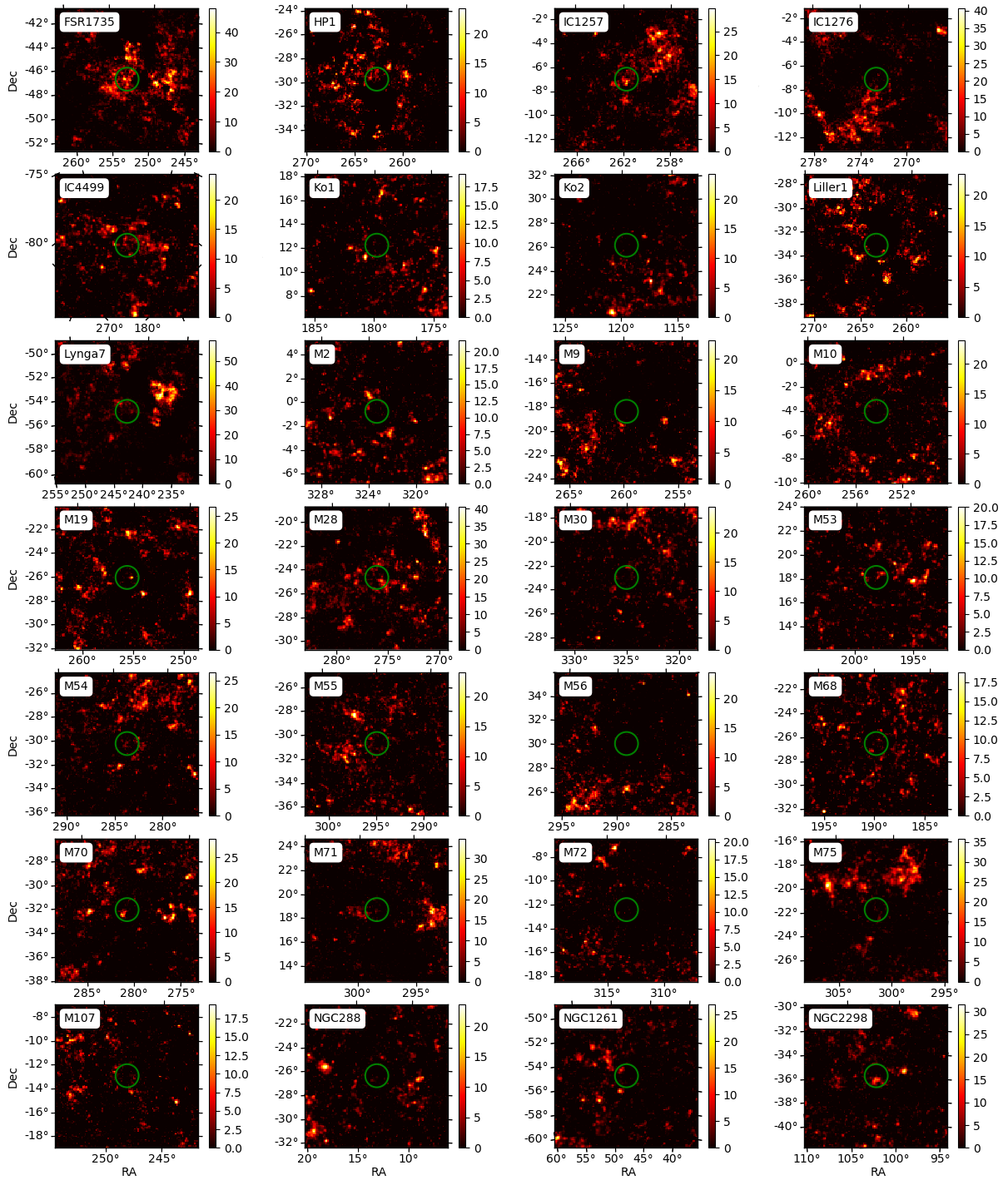}
\caption{TS residuals maps for all Milky Way GCs non-detected in $\gamma$-rays. All maps are centered on the GC positions given by the GLOBCUST catalog. The green circles guide the readers to the center of the maps.}
\label{AllGC}
\end{figure*}

\begin{figure*}
\ContinuedFloat 
\centering
\includegraphics[width=\linewidth]{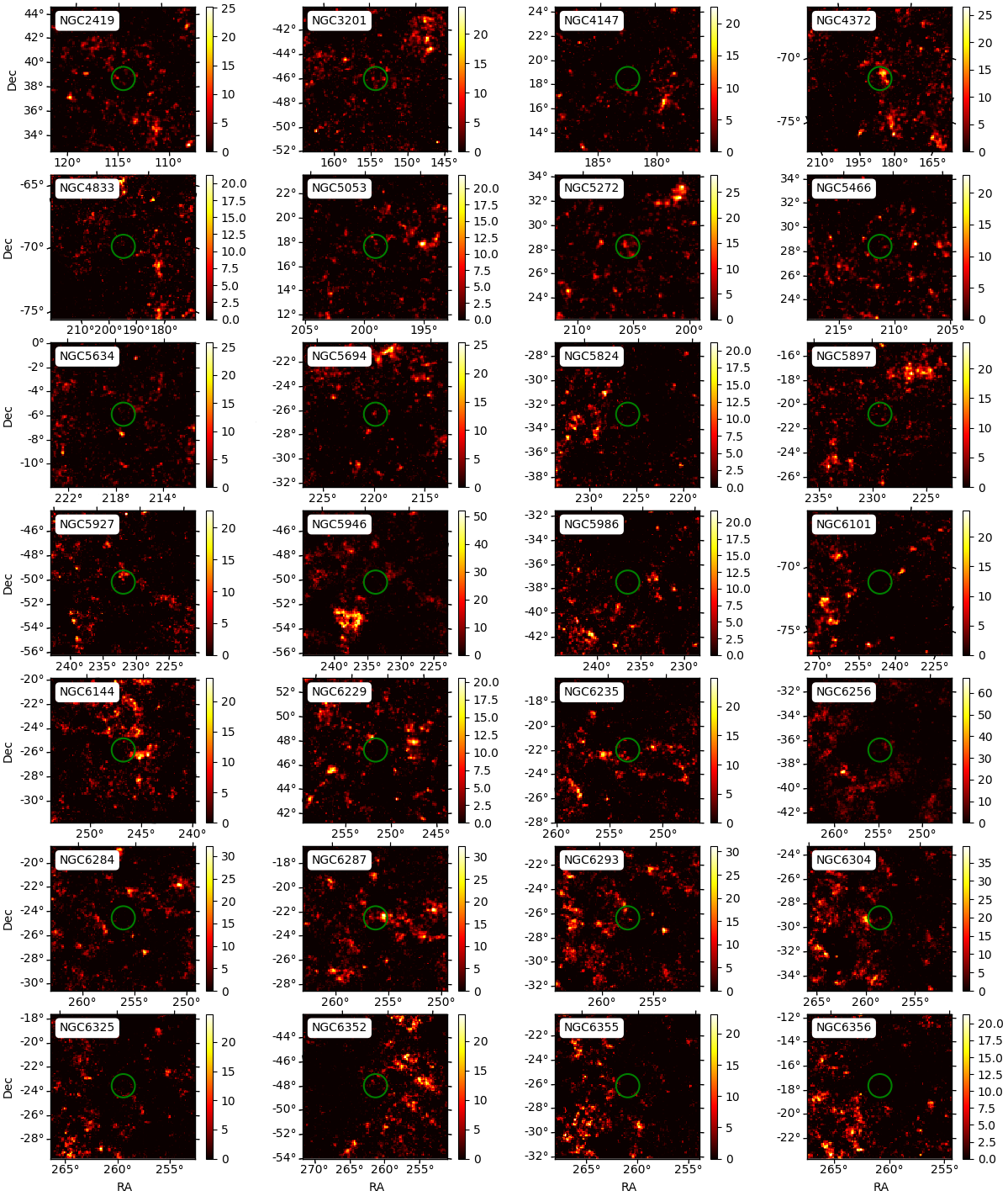}
\caption{TS residuals maps for all Milky Way GCs non-detected in $\gamma$-rays. All maps are centered on the GC positions given by the GLOBCUST catalog. The green circles guide the readers to the center of the maps.}
\label{AllGC}
\end{figure*}

\begin{figure*}
\ContinuedFloat 
\centering
\includegraphics[width=\linewidth]{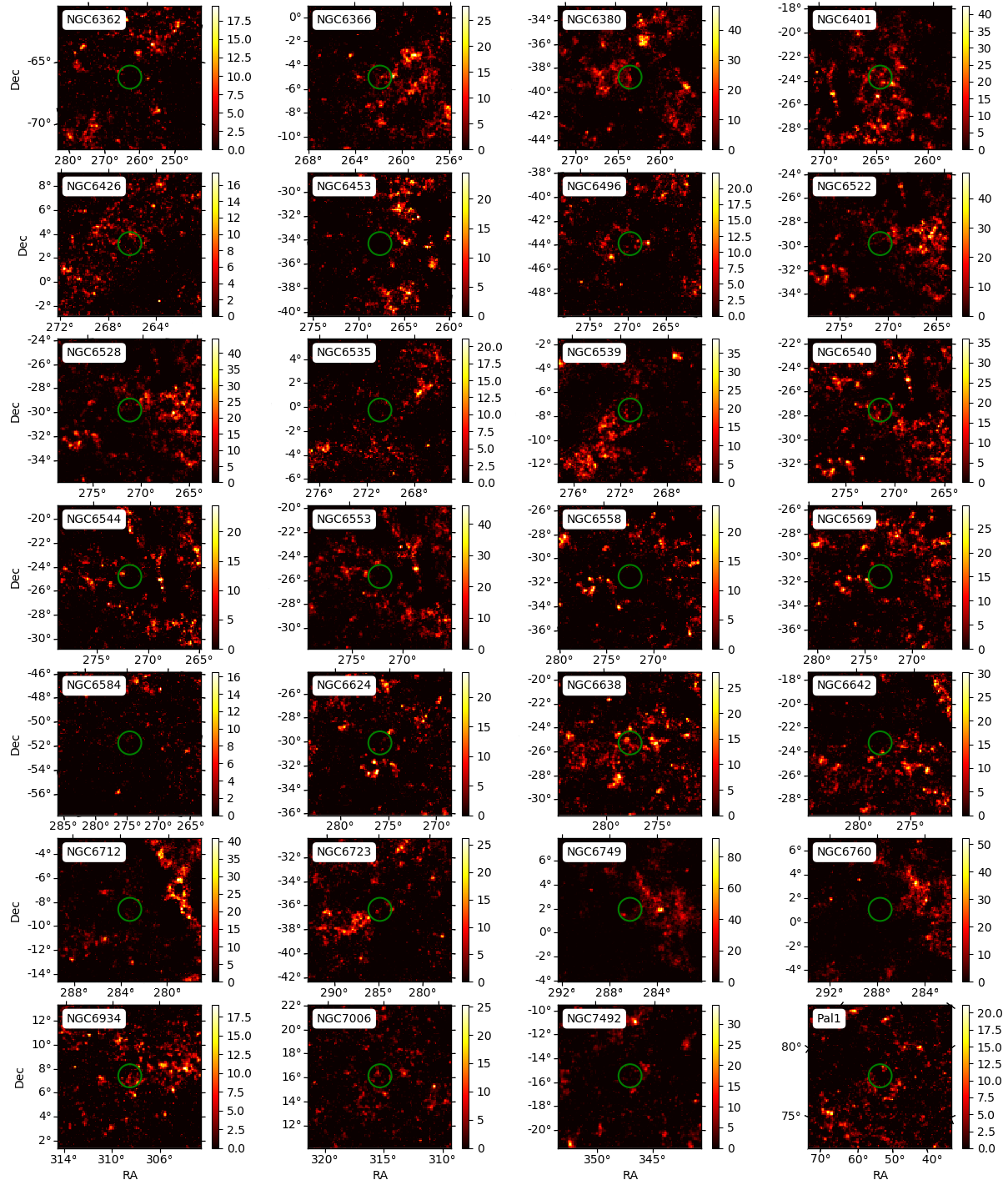}
\caption{TS residuals maps for all Milky Way GCs non-detected in $\gamma$-rays. All maps are centered on the GC positions given by the GLOBCUST catalog. The green circles guide the readers to the center of the maps.}
\label{AllGC}
\end{figure*}

\begin{figure*}
\ContinuedFloat 
\centering
\includegraphics[width=\linewidth]{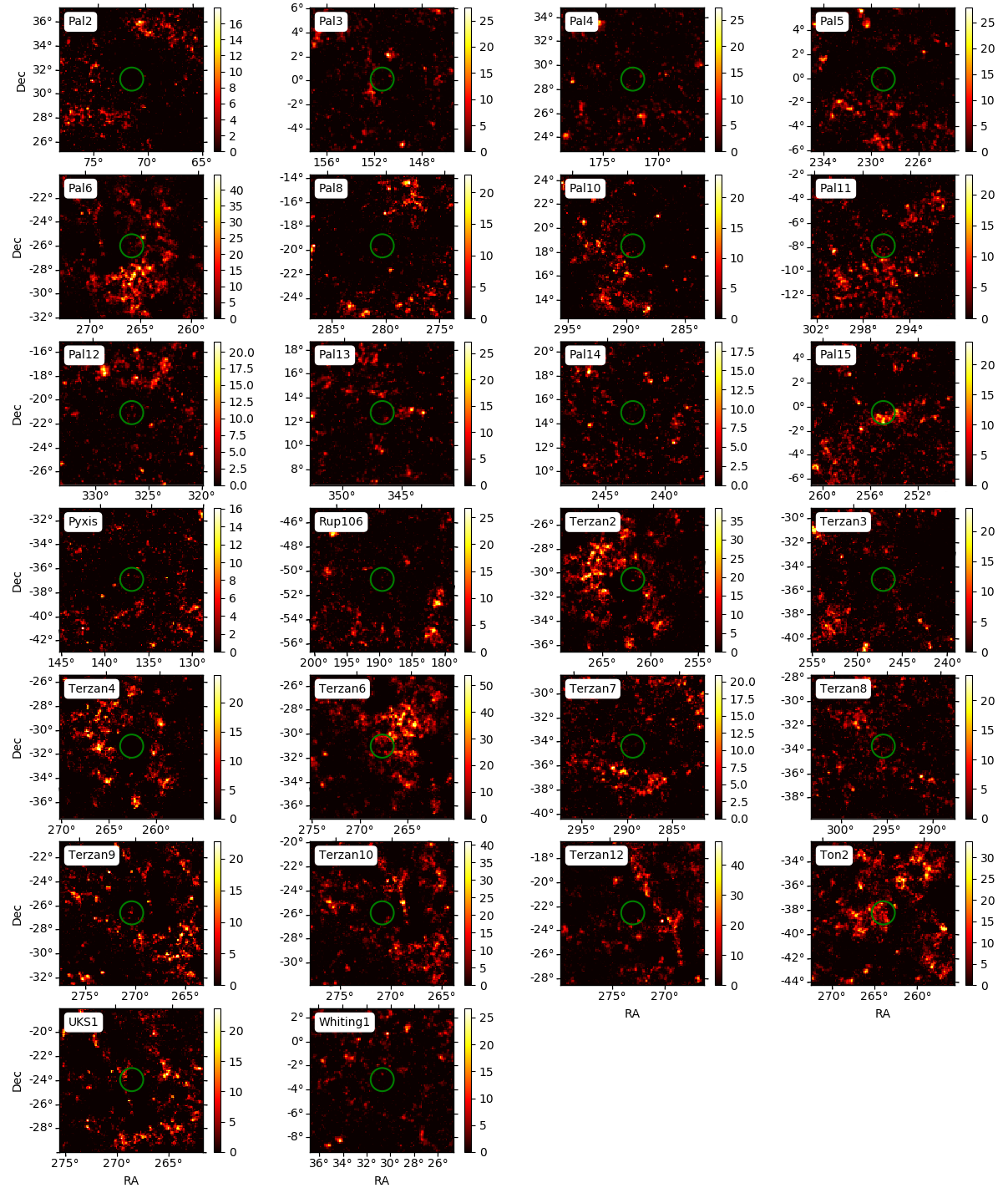}
\caption{TS residuals maps for all Milky Way GCs non-detected in $\gamma$-rays. All maps are centered on the GC positions given by the GLOBCUST catalog. The green circles guide the readers to the center of the maps.}
\label{AllGC}
\end{figure*}

\begin{figure*}
    \centering
    \includegraphics[width=\linewidth]{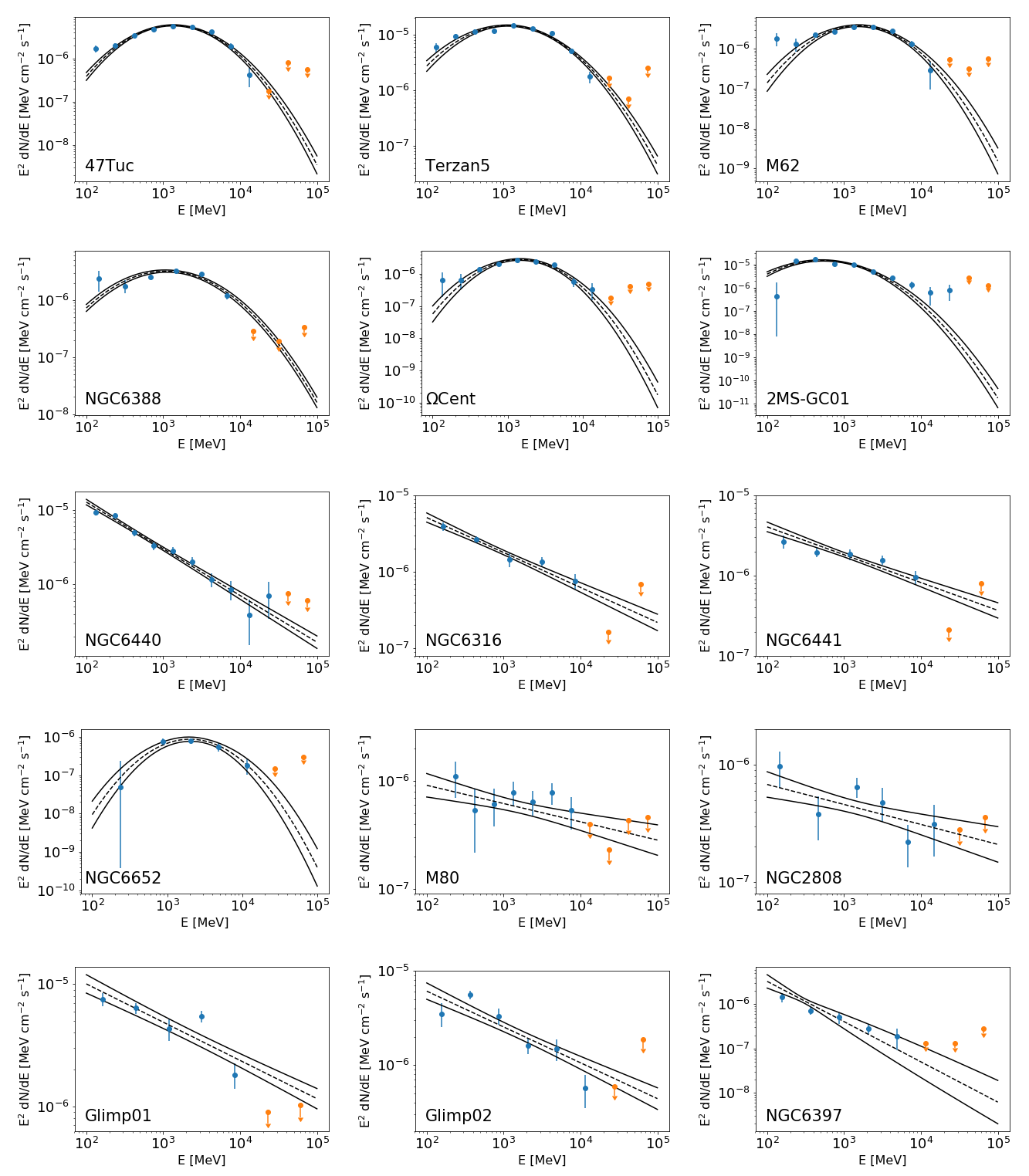}
    \caption{High energy $\gamma$-ray spectra for 19 of the 23 $\gamma$-ray-bright GCs. The panels are organized accordingly with Table \ref{Tabela1}, with higher significance detections located in the top. All spectra are reasonably well fitted with a logparabola or power-law. It is easily noticeable the dominance of logparabola spectral shape between the most significant sources; the low-significance sources, on the other hand, are better described by a power-law. The adopted energy range was from 100 MeV up to 100 GeV.}
    \label{fig:SEDs}
\end{figure*}

\begin{figure*}
    \ContinuedFloat 
    \centering
    \includegraphics[width=\linewidth]{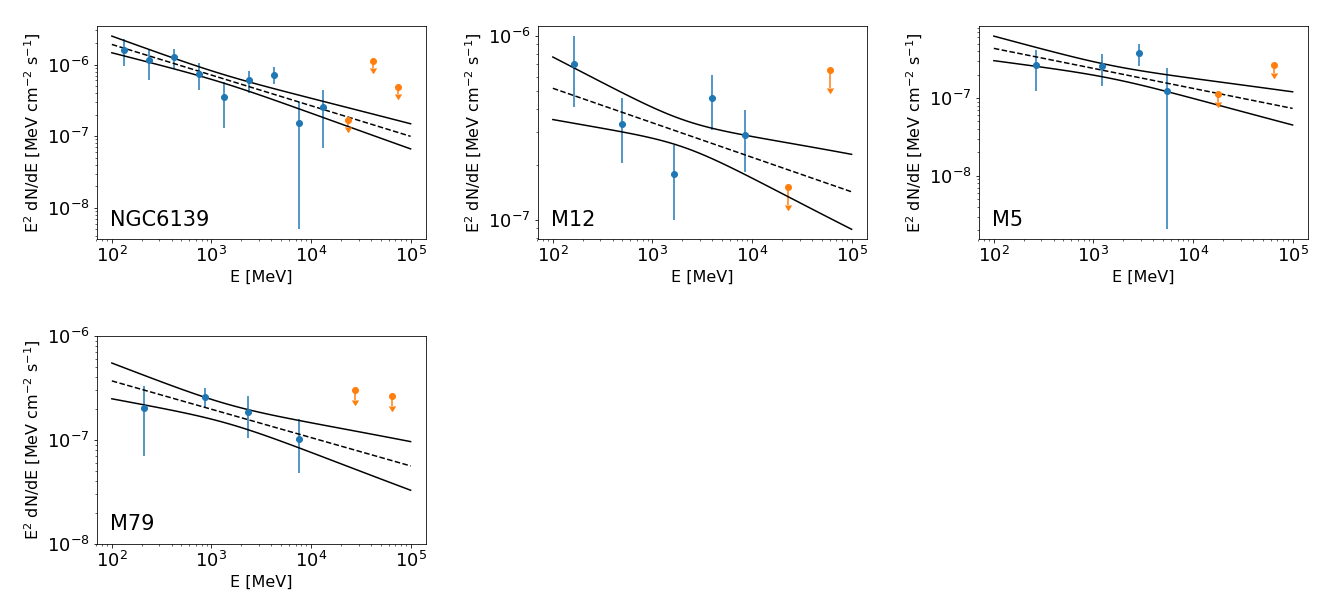}
    \caption{High energy $\gamma$-ray spectra for 19 of the 23 $\gamma$-ray-bright GCs. The panels are organized accordingly with Table \ref{Tabela1}, with higher significance detections located in the top. All spectra are reasonably well fitted with a logparabola or power-law. It is easily noticeable the dominance of logparabola spectral shape between the most significant sources; the low-significance sources, on the other hand, are better described by a power-law. The adopted energy range was from 100 MeV up to 100 GeV.}
    \label{fig:SEDs}
\end{figure*}

\newpage

\begin{table}
\centering
\begin{tabular}{l|c|c}
Cluster & Upper limit & TS \\
        & $10^{-13}$erg cm$^{-2}$s$^{-1}$ \\
\hline           
2MASS-GC02  &  129.6  &  18 \\
AM1  &  32.0  &  0 \\ 
AM4  &  20.3  &  5 \\ 
Arp2  &  8.2  &  1 \\ 
BH176  &  18.2  &  1 \\ 
BH261  &  6.0  &  0 \\ 
Djorg1  &  7.6  &  0 \\ 
Djorg2  &  26.4  &  3 \\ 
E3  &  13.1  &  4 \\ 
Eridanus  &  5.7  &  0 \\ 
ESO452-SC11 & 40.9 & 14 \\
ESO-SC06  &  10.1  &  1 \\ 
FSR1735  &  86.2  &  23 \\ 
HP1  &  35.1  &  1 \\ 
IC1257  &  41.1  &  14 \\ 
IC1276  &  84.1  &  4 \\ 
IC4499  &  22.2  &  7 \\ 
Ko1  &  4.7  &  0 \\ 
Ko2  &  1.7  &  0 \\ 
Liller1  &  8.5  &  0 \\ 
Lynga7  &  107.4  &  3 \\ 
M2  &  6.1  &  0 \\ 
M4  &  54.3  &  24 \\ 
M9  &  1.7  &  0 \\ 
M10  &  5.1  &  0 \\ 
M13  &  32.3  &  23 \\
M15  &  36.4  &  9 \\
M19  &  10.5  &  0 \\
M22  &  48.6  & 23 \\
M28  &  47.1  &  4 \\ 
M30  &  12.8  &  1 \\ 
M53  &  18.7  &  5 \\ 
M54  &  18.8  &  3 \\ 
M55  &  23.2  &  6 \\ 
M56  &  2.2  &  0 \\ 
M68  &  18.6  &  14 \\ 
M69  &  25.2  &  8 \\ 
M70  &  17.2  &  4 \\ 
M71  &  31.6  &  19 \\ 
M72  &  1.5  &  0 \\ 
M75  &  11.5  &  1 \\ 
M92  &  32.3  &  18 \\ 
M107  &  6.3  &  0 \\ 
NGC288  &  5.7  &  0 \\ 
NGC362  &  19.2  &  16 \\ 
NGC1261  &  10.4  &  7 \\ 
NGC1851  &  20.3  &  17 \\ 
NGC2298  &  28.8  &  14 \\ 
NGC2419  &  10.9  &  7 \\ 
NGC3201  &  17.4  &  3 \\ 
NGC4147  &  4.9  &  0 \\ 
NGC4372  &  35.1  &  10 \\ 
NGC4833  &  4.0  &  0 \\ 
NGC5053  &  9.5  &  1 \\ 
NGC5272  &  15.2  &  8 \\ 
NGC5286  &  35.6  &  15 \\ 
NGC5466  &  6.0  &  0 \\ 
NGC5634  &  3.6  &  0 \\ 
NGC5694  &  14.2  &  5 \\ 
NGC5824  &  3.3  &  0 \\ 
NGC5897  &  8.2  &  0 \\ 
NGC5927  &  12.5  &  0 \\ 
NGC5946  &  8.0  &  0 \\ 
NGC5986  &  3.7  &  0 \\ 
NGC6101  &  3.3  &  0 \\ 
NGC6144  &  6.3  &  0 \\ 
NGC6229  &  1.8  &  0 \\

\end{tabular}

\caption{ Energy flux $2\sigma$ upper limits and TS values for all 129 GCs non-detected in $\gamma$-rays. The energy range adopted is from 100 MeV up to 100 GeV.}
\label{Tabela2_1}
\end{table}

\begin{table}
\centering
\begin{tabular}{l|c|c}
Cluster & Upper limit & TS \\
        & $10^{-13}$erg cm$^{-2}$s$^{-1}$ \\
\hline          
NGC6235  &  4.9  &  0 \\ 
NGC6256  &  45.9  &  1 \\
NGC6284  &  2.1  &  0 \\ 
NGC6287  &  43.2  &  12 \\ 
NGC6293  &  15.7  &  1 \\ 
NGC6304  &  32.9  &  6 \\ 
NGC6325  &  7.9  &  0 \\ 
NGC6342  &  32.3  &  21 \\ 
NGC6352  &  6.4  &  0 \\ 
NGC6355  &  2.6  &  0 \\ 
NGC6356  &  2.2  &  0 \\ 
NGC6362  &  2.5  &  0 \\ 
NGC6366  &  20.8  &  2 \\ 
NGC6380  &  84.9  &  21 \\ 
NGC6401  &  128.4  &  11 \\ 
NGC6426  &  4.3  &  0 \\ 
NGC6453  &  3.0  &  0 \\ 
NGC6496  &  3.2  &  0 \\ 
NGC6517  &  34.6  &  6 \\ 
NGC6522  &  53.3  &  1 \\ 
NGC6528  &  54.1  &  3 \\ 
NGC6535  &  5.8  &  0 \\ 
NGC6539  &  11.3  &  0 \\ 
NGC6540  &  60.4  &  5 \\ 
NGC6544  &  12.1  &  0 \\ 
NGC6553  &  29.3  &  4 \\ 
NGC6558  &  2.2  &  0 \\ 
NGC6569  &  11.4  &  0 \\ 
NGC6584  &  2.3  &  0 \\ 
NGC6624  &  137.3  &  4 \\ 
NGC6638  &  48.9  &  5 \\ 
NGC6642  &  8.8  &  0 \\ 
NGC6712  &  17.8  &  6 \\ 
NGC6723  &  13.6  &  1 \\ 
NGC6749  &  113.9  &  0 \\ 
NGC6760  &  111.9  &  1 \\ 
NGC6934  &  17.7  &  7 \\ 
NGC7006  &  20.3  &  4 \\ 
NGC7492  &  9.5  &  1 \\ 
Pal1  &  13.7  &  4 \\
Pal2  &  2.1  &  0 \\ 
Pal3  &  8.5  &  1 \\ 
Pal4  &  2.8  &  0 \\ 
Pal5  &  11.2  &  0 \\
Pal6  &  3.00  &  5 \\
Pal8  &  11.5  &  0 \\ 
Pal10  &  3.1  &  0 \\ 
Pal11  &  12.7  &  2 \\ 
Pal12  &  4.8  &  0 \\ 
Pal13  &  15.9  &  5 \\ 
Pal14  &  3.0  &  0 \\ 
Pal15  &  15.1  &  5 \\ 
Pyxis  &  3.3  &  0 \\ 
Rup106  &  13.9  &  2 \\ 
Terzan1  &  99.7  &  18 \\ 
Terzan2  &  42.2  &  8 \\ 
Terzan3  &  3.4  &  0 \\ 
Terzan4  &  2.9  &  0 \\ 
Terzan6  &  129.7  &  14 \\ 
Terzan7  &  8.4  &  0 \\ 
Terzan8  &  21.1  &  4 \\ 
Terzan9  &  40.8  &  5 \\ 
Terzan10  &  34.2  &  5 \\ 
Terzan12  &  3.4  &  0 \\ 
Ton2  &  99.9  &  12 \\ 
UKS1  &  7.3  &  0 \\ 
Whiting1  &  2.0  &  0 \\ 
\end{tabular}

\caption{ Continuation of table A1.}
\label{Tabela2_2}
\end{table}

\bsp	
\label{lastpage}
\end{document}